# ON THE COMPARISON OF MODELS AND EXPERIMENTS IN THE STUDY OF DNA OPEN STATES: THE PROBLEM OF DEGREES OF FREEDOM

## AUTHORS

Alexey S. Shigaev[1,*], and Victor D. Lakhno[1]

[1]The Laboratory of Quantum-Mechanical Systems, the Institute of Mathematical Problems of Biology, Russian Academy of Sciences – A Branch of the KIAM RAS, Pushchino, Moscow Region, 142290, Russia
* To whom correspondence should be addressed. Email: shials@rambler.ru

## GRAPHICAL ABSTRACT

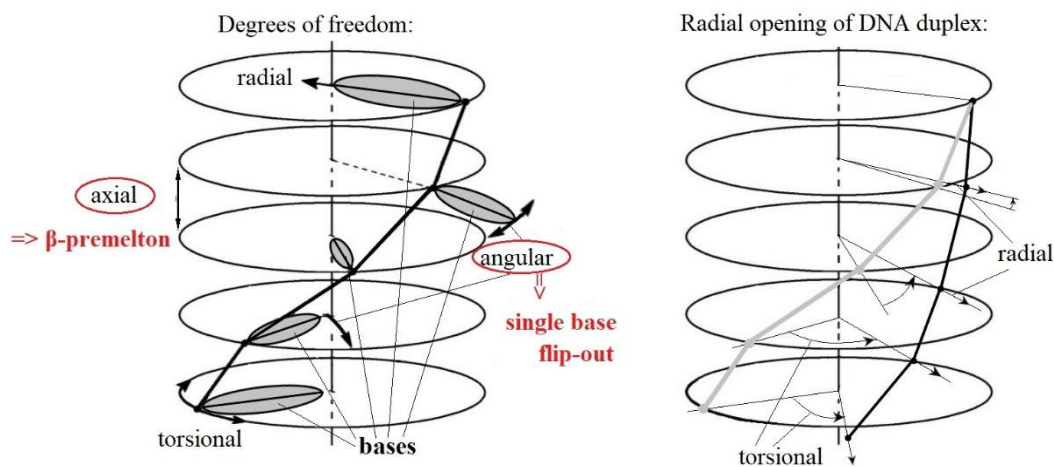

## ABSTRACT

Simple mechanical models of DNA play an important role in studying the dynamics of its open states. The main requirement when developing a DNA model is the correct selection of its effective potentials and parameters based on experimental data. At the same time, various experiments allow us to "see" different types of DNA open states. Consideration of this feature is one of the most important conditions in the development, optimization, and parameterization of any mechanical model. Violation of this condition, i.e., the comparison of incomparable characteristics, leads to critical errors. The present investigation is devoted to the problem of degrees of freedom of DNA bases taken into account in mechanical models. Using the Peyrard-Bishop-Dauxois model as an example, two types of errors in interpreting experimental data when compared with the model are examined. The first one is a mismatch between the open state types in the model and experiment. The second one is an incorrect specification of the "threshold coordinate" of the open state. The concept of the effective total threshold coordinate of the radial separation of DNA strands for registration of opening is introduced. It is shown that correct interpretation of experimental data can actually eliminate discrepancies with theory.

**INTRODUCTION**

Local transient openings of the DNA double helix play an important role in the processes of reading the genetic code. DNA open states occur as a result of complete or partial breaking of complementary H-bonds in one or more adjacent nucleotide pairs; a key feature of any DNA open state is the accessibility of imino protons involved in the formation of complementary H-bonds to solution molecules [1]. Different degrees of freedom of DNA bases are involved in the formation of different open states [1].

The so-called mechanical models are one of the most important tools for studying the dynamics of DNA open states. Most mechanical models use one to four variables to describe each DNA base pair, see [1, 2]. The variables in models of this type correspond to certain degrees of freedom of DNA bases. Accordingly, one can argue the specification of different mechanical models of DNA for the study of different open states.

A key feature of DNA mechanical models is their comparatively low computational resource consumption compared to full-atom molecular dynamics. The smallest time intervals between the openings of the Watson-Crick DNA helix are on the millisecond timescale [3]. Therefore, full-atom MD simulations are unlikely to provide a complete picture of the dynamics of DNA opening.

On the other hand, a simplified description of a DNA duplex inevitably limits the possibilities of theoretical research. Open states of DNA differ in the contribution of different degrees of freedom of the bases at the beginning of the molecular dynamics trajectory of opening. Therefore, even with a successful choice of potentials, any given mechanical model of DNA is not suitable for studying all types of open states. In other words, universal mechanical models of DNA simply do not exist. Moreover, even with the correct choice of potentials, there is a risk of parameterization errors, which can render the results completely useless. One way to avoid such errors is to strive for an exact correspondence between the types of DNA open states in the model and in the experiment.

The present investigation examines the correspondence between the degrees of freedom of bases during the formation of DNA open states in a mechanical model and experiments. The Peyrard-Bishop-Dauxois (PBD) model [4] is used as an example. Accordingly, the radial elongation of the complementary H-bonds – that is, the radial degree of freedom of base pairs – was chosen as the key degree of freedom. The relationship between this degree of freedom and the torsional stresses of the sugar-phosphate backbone is examined. The concept of local cooperativity during radial-torsional opening of the DNA duplex is described.

An analysis of DNA base displacements during the formation of various open states was carried out. One of the fundamental errors of researchers is demonstrated – the comparison of the kinetics of concerted radial openings of the DNA duplex in the PBD model with single angular openings of the bases in experiments. The introduction of an effective total threshold coordinate of the radial displacement of the bases instead of the displacement thresholds of single bases is justified. It is shown that the introduction of the total radial threshold coordinate dramatically reduces the discrepancies between the results of the PBD model study and the experimental data.

## 1. RELATIONSHIP BETWEEN THE DEGREES OF FREEDOM OF DNA BASES. THE PROBLEM OF LOCAL COOPERATIVITY.

Let us consider the problem of the relationship between the degrees of freedom of DNA base pairs. Figure 1 shows the main degrees of freedom for the bases in most simple mechanical models of DNA. These degrees of freedom are sufficient for a simplified understanding of the "stereometry" of DNA denaturation

The Roman numeral I denotes the radial elongation of the bundle of complementary H-bonds in a nucleotide pair. Many simple mechanical models of DNA are limited to a radial degree of freedom for each base pair, see [4, 5, 6, 7]. Hence, they use a single variable — the radial one, accordingly. The Roman numeral II in Fig. 1 corresponds to the torsional (or helical) degree of freedom. This degree of freedom usually corresponds to the angular variable in "combined" mechanical models of DNA, see [8, 9 , 10, 11, 12, 13, 14, 15]. Typically, models of this type include a combination of a radial degree of freedom (I) with a torsional one (II).

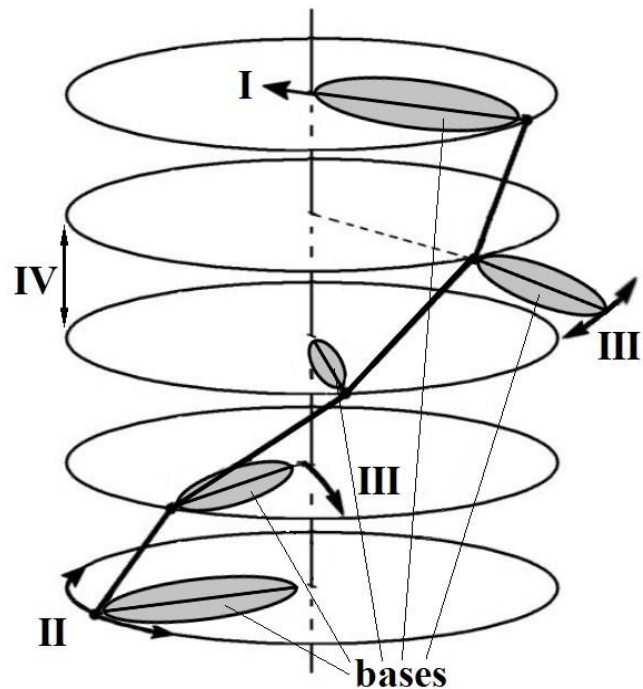

**Figure 1:** Simplified schematic representation of the DNA bases' main degrees of freedom. Only one strand of the duplex is shown. (I) radial degree of freedom – elongation of the bundle of complementary H-bonds in a base pair; (II) torsional (helical) degree of freedom – change in the number of bases per complete turn of the Watson-Crick helix; (III) angular degree of freedom – rotation of a single base around the axis of the sugar-phosphate backbone; (IV) axial degree of freedom – elongation of the duplex along the longitudinal axis.

Negative displacements are greatly hindered by steric restrictions for degrees of freedom (I) and (IV): this fact is generally taken into account in the DNA models.

The angular degree of freedom of the bases is indicated by the Roman numeral III. This degree of freedom corresponds to rotation of the base around the site of its covalent attachment to the sugar-phosphate backbone. It is shown twice in Figure 1 – specifically to clearly illustrate the result of the angular escape of the base from the Watson-Crick helix, the so-called flipping (or flip-out) [16, 17, 18]. The degree of freedom (III) is relatively weakly coupled to the other degrees of freedom. Therefore, base flip-outs are generally non-cooperative [19, 20, 21, 22]. It is worth noting that the "illustrative" term "flip-out" began to be used only after the well-known crystallographic studies of S. Klimasauskas et al. in 1994 in 1994 [23].

The Roman numeral IV denotes the axial degree of freedom – the elongation of DNA along its longitudinal axis. This degree of freedom is also present in some combined models [14]. This is the only conformational change in DNA that does not necessarily lead to the breakage of complementary H-bonds, see below.

Let us consider the relationship between the radial (I) and torsional (II) degrees of freedom of the bases during the radial opening of a DNA duplex. The onset of the radial opening is shown in Figure 2.

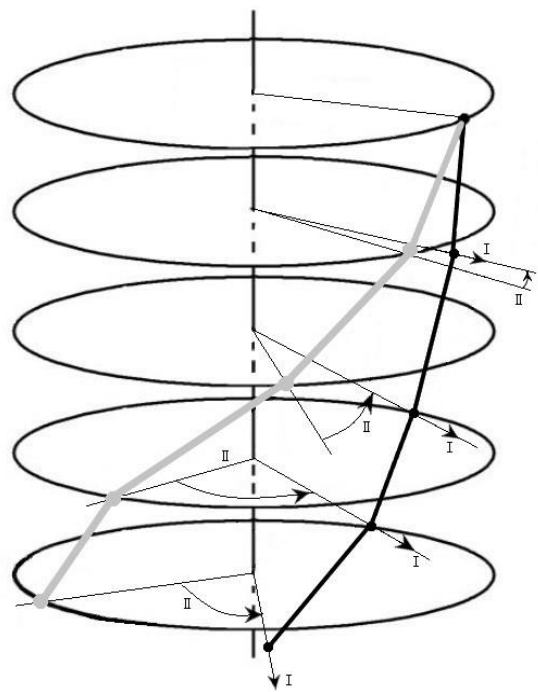

**Figure 2:** The onset of the radial opening of a region in DNA duplex. The initial positions of the attachment sites of the bases to the sugar-phosphate backbone are shown as gray dots. The initial positions of the backbone regions connecting the bases are shown as gray lines. The bases themselves are not shown. The new positions of the bases and fragments of the sugar phosphate backbone are shown as solid dots and bold lines, respectively. Numbers (I) and (II) indicate the degrees of freedom involved; the angles between the initial and final directions of the bases are also shown.

Radial elongation of the complementary H-bonds bundle inevitably causes a stress on the bond angles of the DNA sugar-phosphate backbone. Clearly, the stress is relieved by local unwinding of the Watson-Crick helix. The higher the rigidity of the valence angles of the sugar-phosphate backbone, the more effective the relationship between the degrees of freedom (I) and (II).

This relationship prevents the radial opening of an individual DNA base pair – this is precisely the main feature that differs radial opening (degrees of freedom I+II) from angular opening (degree of freedom III) in terms of cooperativity. The rigidity of the sugar-phosphate backbone should lead to the redistribution of the energy of any radial fluctuations over several adjacent base pairs.

In other words, radial displacements of base pairs are always coupled through torsional stresses. Hereinafter, we will refer to this effect as local cooperativity. The concept of local cooperativity of DNA denaturation was first introduced by V. Ivanov et al. in 2005 within the framework of a two-state approach [24]. In the present study, we aim to demonstrate the mechanism of local cooperativity by considering mechanical models of DNA.

The spatial extent of local cooperativity can be estimated from some experimental data. For example, experiments by G. Zocchi's group revealed the so-called "critical length" phenomenon of DNA oligomers [25]. Many of the DNA oligomers studied had the 5'-$G_nA_mG_n$-3' primary structure, where A is the AT pair and G is the GC pair [25, 26, 27, 28]. We do not provide details of the primary structure for simplicity: AT ≡ TA, GC ≡ CG.

In the case of $2n + m > 20$, a specific technique made it possible to register so-called "melting intermediates" – oligomers with a denatured AT-rich center. Oligomers with the property $2n + m \leq 20$ dissociated without forming an intermediate, like simple chemical complexes. At the same time, DNA with the structure 5'-$G_nA_m$-3' did not melt "at once" even with a short chain length ($n + m < 20$), but formed intermediates [25].

The phenomenon of critical length complements the picture of local cooperativity. High-amplitude radial fluctuations occur predominantly in AT-rich domains. The interplay between the radial (I) and torsional (II) degrees of freedom during opening causes local unwinding – an increase in the total number of base pairs per full turn of the Watson-Crick helix. The unwinding of the middle domain should create torsional stresses in adjacent regions. The stresses are caused by the rigidity of the covalent bonds of the sugar-phosphate backbone. Nevertheless, the resulting stress is easily released at the ends of the chain in the case of short DNA.

The denaturation behavior of an oligomer with the 5'-$G_nA_mG_n$-3' structure is shown in Figure 3. Clearly, the spatial extent of local cooperativity effects in such a structure exceeds the distance between the edge of the nascent radial opening and the chain ends at $2n + m \leq 20$. Therefore, the despiralization (torsional) stress is released at both ends.

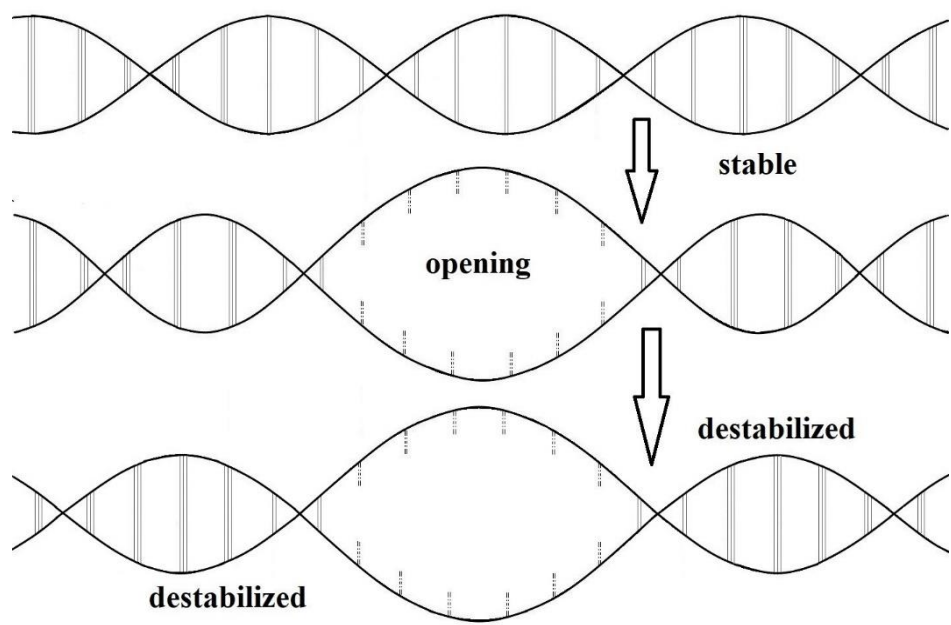

**Figure 3:** See text for explanation.

The despiralization should persist for some time after the stress is released. The stress-driven DNA strand separation is a well-known phenomenon, see [29, 30, 31, 32]. Therefore, DNA with the structure 5'-$G_nA_mG_n$-3', where $2n + m \leq 20$, unwinds without the formation of intermediates.

The situation is fundamentally different for denaturation of DNA with the 5'-$G_nA_m$-3' structure. Any stress is released at the side with the open AT-rich terminal domain. Therefore, the critical length phenomenon has not been observed for two-domain oligomers [25]. The corresponding behavior of DNA with a 5'-$G_nA_m$-3' structure is shown in Fig. 4.

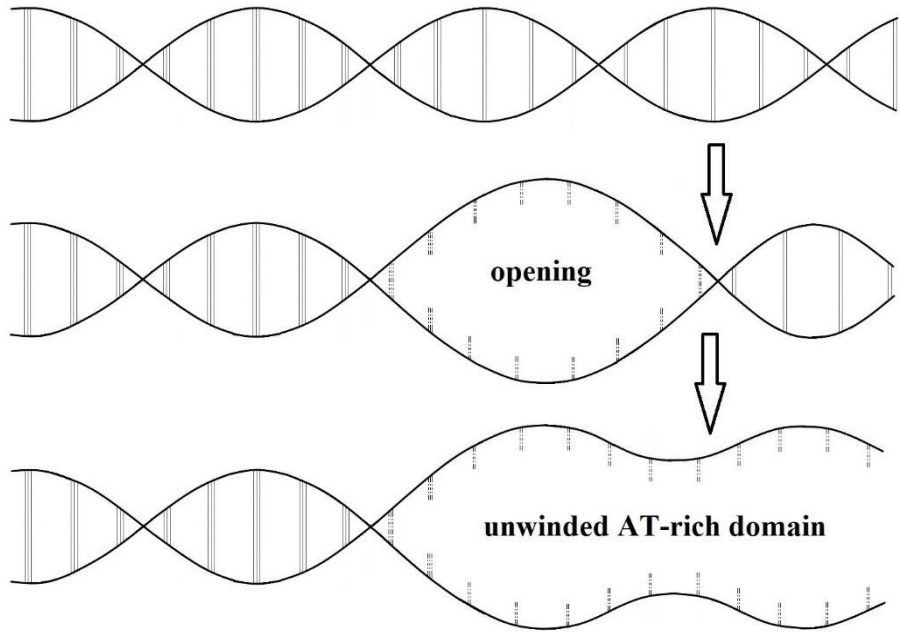

**Figure 4:** See text for explanation.

This effect can literally be "felt with your hands." It is enough to twist any rope into a double spiral on your own and then stretch it "across" in different places. Apparently, the extent of local cooperativity effects is comparable to the length of a full turn of the Watson-Crick helix: about 10 base pairs.

There are three key points that require further clarification regarding the relationship between the degrees of freedom of DNA bases. First, torsional fluctuations can also lead to radial opening of DNA. In other words, the degree of freedom (II) may well play a key role in the radial opening of the duplex. We have not emphasized this detail above, since the concerted, cooperative radial opening of nucleotide pairs in this case is obvious.

Second, the degree of freedom (II) is closely related to the degree of freedom (IV). Torsional stresses during duplex unwinding can lead both to radial separation of DNA strands, and to breaks of stacking interactions between adjacent base pairs. This correspondent conformational change is clearly shown in Figure 5.

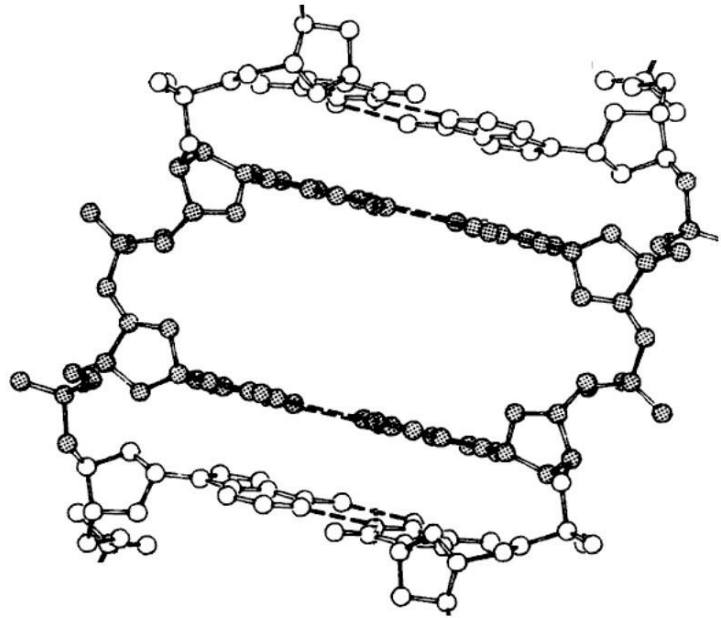

**Figure 5:** A local change in DNA conformation – β-premelton. The nucleotide pairs between which stacking interactions are disrupted are shaded in gray.

Such conformational changes in DNA were first described in 1983 as β-premeltons [33]. Later, the possibility of a conformational transition of DNA to an elongated form (so-called S-DNA) was demonstrated in micromechanical experiments [34]. The absence of DNA unwinding when it is elongated by 1.7 times clearly indicates that the complementary binding of the chains is intact. Clearly, the integrity of complementary H-bonds is not disrupted in this case. Therefore, β-premelton cannot be considered as an open state of DNA (see [1] for definition "DNA open state"). At the same time, this conformational transition obviously has some potential energy – as we will see below, this is very important!

Third, it should be emphasized once again that the angular degree of freedom of the bases (III) is almost not related with all the others. Therefore, the angular exits of the bases from the Watson-Crick helix (the flip-outs) occur non-cooperatively [19, 20, 21, 22]. The absence of a direct relation between the angular degree of freedom (III) and the torsional one (II) is also shown in silico [17, 35, 36]. An example of the angular exit of a base from the Watson-Crick helix is shown in Figure 6. It is clearly seen that if the changes affect only the degree of freedom (III), no torsional changes in the sugar-phosphate backbone occur.

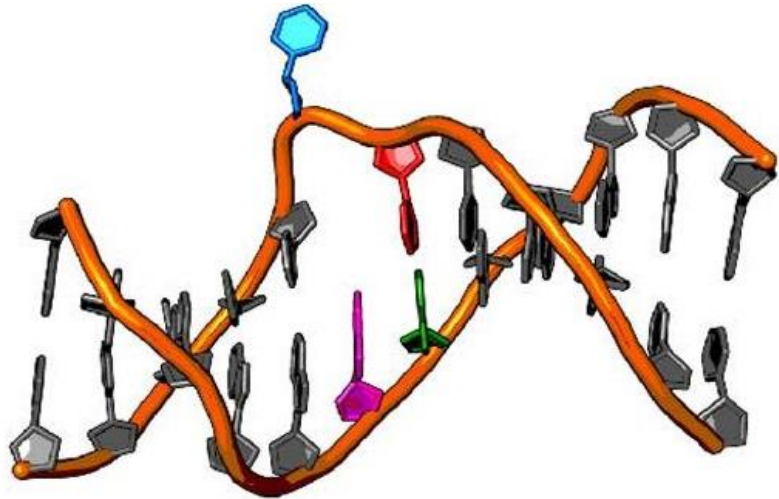

**Figure 6:** Flip-out of the thymine base (blue) from a Watson-Crick helix [18].

In fact, all experimental data on flip-out of DNA bases from the Watson-Crick helix (as of the end of 2012) prove that openings of this type occur strictly one at a time, see Chapter 5 of Review [1]. The only exceptions are paired flip-outs in the so-called A-tracts – special structures with three-center complementary H-bonds [37, 38, 39, 40].

At the same time, the central degree of freedom in many mechanical models of DNA is precisely the angular displacement of the bases [41, 42, 43, 44, 45, 46]. Angular displacements in these models are quite cooperative. However, there are no contradictions between the models and experiments in this case: apparently, the angular displacements of the bases are quite capable of moving along the DNA duplex provided that changes in the degree of freedom (III) have a moderate amplitude. This is how one of the mechanisms for the transfer and localization of vibrational energy of atomic groups is realized. If a certain threshold amplitude is exceeded, the complementary H-bonds in a base pair are replaced by bonds with water molecules — that is, an open state is formed.  Estimates of the corresponding amplitudes and energies are given in Section 6.4 of [1].

So, radial-torsional openings of the DNA duplex are associated with cooperative breakage of complementary H-bonds in several adjacent base pairs due to the close relationship between the degrees of freedom (I) and (II). In contrast, flip-out typically affects only one base due to the weak coupling of the angular degree of freedom (III) with the others.

At the same time, the richest experimental data has been accumulated specifically on DNA base flip-outs, see [1]. This leads to misunderstandings when comparing the behavior of models and experiments. It is precisely the experimental data on flip-outs that are used as the basis for criticizing the mechanical models of DNA, whose variables correspond to degrees of freedom (I) and (II). Statements such as "...the PBD model is extremely attractive to physicists since it requires little knowledge about DNA biophysics..." and even criticism of the mechanical models of DNA as an entire field of research are common [47].

The criticism seems quite convincing – but only for those researchers who do not consider that different degrees of freedom of the bases are involved in the formation of different types of DNA open state. The experimental equilibrium constants $K_d$ for the flip-out reaction are indeed very small: they are approximately $10^{-5}$ for AT pairs and approximately $10^{-6}$ for GC pairs [19, 20, 21, 22]. Similar constants for single radial openings of base pairs in the PBD model are approximately 0.016 and 0.004, respectively [48].

However, this difference of almost three orders of magnitude is due to the fact that critics simply compare the incomparable. This is easily demonstrated within the framework of the concept of the effective radial coordinate, which will be described below. .

## 2. EFFECTIVE THRESHOLD COORDINATE AND ITS USE TO DESCRIBE THE RADIAL SEPARATION OF DNA STRANDS

Let us consider the problem of comparing theory and experiment using the PBD model as an example. The state of each base pair in the PBD model is described by a single variable – the radial elongation of the bundle of complementary H-bonds in the base pair. This variable corresponds to the radial degree of freedom (I), see Fig. 1. The Hamiltonian of the PBD model has the form [4]:

$$H = \sum \left[ \frac{m\dot{y}_n^2}{2} + V(y_n) + W(y_n, y_{n-1}) \right] \quad (1)$$

The first term is the kinetic energy of the base pair; $m$ is the effective mass of the model site corresponding to the base pair; coordinate $y_n$ is the radial elongation of the complementary bundle of H-bonds divided by $\sqrt{2}$, see [6].

The second term, $V(y_n)$, corresponds to the on-site potential:

$$V(y_n) = D_n \left( \exp[-a_n y_n] - 1 \right)^2 \quad (2)$$

where $D_n$ is the depth of the potential well, and $a_n$ is its inverse width correspondingly. The third term is the inter-site potential $W(y_n, y_{n-1})$:

$$W(y_n, y_{n-1}) = \frac{k}{2} \left( 1 + \rho \cdot \exp\left[ -\alpha (y_n + y_{n-1}) \right] \right) \cdot (y_n - y_{n-1})^2 \quad (3)$$

where $k$ is the elastic constant describing stacking interactions, and $\alpha$ is the decay constant, indirectly accounting for the finiteness of stacking interactions. The dimensionless cooperativity parameter $\rho$ phenomenologically takes into account the nonlinear nature of stacking interactions in DNA [4].

The radial extension of the bundle of complementary H-bonds is the only degree of freedom of a base pair in the PBD model. The coordinate $y_n$ in (1)-(3) corresponds to the degree of freedom (I) in Figure 1. The torsional degree of freedom (II) is absent in the PBD model. Hence, direct consideration of the relationship between degrees of freedom (I) and (II) in the PBD model is impossible. This consideration is possible only indirectly. For example, the extent of local cooperativity in the PBD model is regulated through the parameters of the inter-site potential (2).

Any model that takes into account the radial degree of freedom (I) for a nucleotide pair as the only one has similar properties. These properties impose additional requirements on the interpretation of experiments when comparing them with models similar to the PBD model. Any simple mechanical model of DNA considers the duplex without taking solvent into account. At the same time, the accessibility of the bases to solution molecules is a key criterion of the open state [1]. Therefore, theorists have to define artificially specific values for the radial coordinate $y_n$ – thresholds $y_{thres}$ ("threshold value"). The base pair is conventionally considered open when $y_n > y_{thres}$, regardless of whether the $y_{thres}$ value is sufficient for solvent molecules to enter between the duplex strands.

Ambiguity of terminology and confusion in degrees of freedom are the main reasons for the misinterpretation of many experiments when comparing them with the behavior of models. The radial coordinate $y_n$ in the PBD model corresponds to the elongation of the complementary H-bonds bundle divided by $\sqrt{2}$, see [6]. The value $y_{thres}$ = 2.1 Å was used in early stochastic studies of DNA behavior at moderate temperatures (T = 301 K) [49, 50]. This seems quite logically justified: $2.1 \cdot \sqrt{2} \approx 3$ Å, it corresponds to the size of a water molecule. That is, the DNA strands must separate enough to accept an $H_2O$ molecule. Therefore, this value of $y_{thres}$ is the lower border of the threshold for the DNA open state.

Such $y_{thres}$ values are considered normal in studies of the PBD model and the so-called ePBD – the extended version with a heterogeneous value of elastic constant $k$ in intersite potential (2) [51]. Moreover, many researchers have used significantly smaller values of the threshold coordinate: $y_{thres}$ = 1.5 Å [52, 53], $y_{thres}$ = 1 Å [51], and even $y_{thres} \leq 0.24$ Å [54, 55]. In the investigation by S. Ares and G. Kalosakas cited above, $K_d$ constants of the order of $10^{-2} - 10^{-3}$ were obtained for $y_{thres}$ = 1.5 Å [48].

A simple analysis can prove that these $y_{thres}$ values are too small to be compared with real experiments when it comes to single-base openings. Therefore, the obtained kinetics of the open states cannot be directly compared with experiments. This is true precisely for the angular escape of single bases (flip-out), with the kinetics of which comparisons are often carried out, see [56].

Indeed, for example, in the AT pair, the exchangeable proton of the imino group is located on the thymine residue, which has a linear size of approximately 3.5 Å. The exit angle of the base during flip-out is approximately 180° [17] (see also Figure 6). Accordingly, the length of the imino proton trajectory is $3.5 \pi = 11$ Å, i.e. $3.5 \pi \cdot \sqrt{2} \approx 7.5$ units of the radial coordinate. This is 4-5 times greater than the $y_{thres}$ values given in theoretical works. For guanine base, similar calculations yield over 13 units!

The results of comparison of theory and experiments turn out to be absurd for such high values of the "real" threshold. Most studies of the PBD model were carried out with the set of parameters obtained in 1998 by A. Campa and A. Giansanti: $k$ = 0.025 eV Å$^{-2}$, $\rho$ = 2, $\alpha$ = 0.35 Å$^{-1}$, $D_{AT}$ = 0.05 eV, $D_{GC}$ = 0.075 eV, $a_{AT}$ = 4.2 Å$^{-1}$, $a_{GC}$ = 6.9 Å$^{-1}$ [57]. Let us define the opening of a single AT pair as the condition for the coordinates of its neighbors to be equal to zero: $y_n$ = $y_{thres}$, $y_{n-1}$ = $y_{n+1}$ = 0. Next, we substitute $k$ = 0.025 eV Å$^{-2}$ and $D_{AT}$ = 0.05 eV into the expression for the constant $K_d$ for the equilibrium case [58]:

$$K_d = \frac{1}{Z}\exp\left[-\frac{D_{AT} + k(1+\rho\cdot\exp[-\alpha\cdot y_n])y_n^2}{k_B T}\right] \qquad (4)$$

where $k_B$ is the Boltzmann constant and $Z$ is the partition function corresponding to the PBD model. It is approximately $Z$ = 7.8 for the AT pair at 300 K [59], which gives an estimate of $K_d \approx 0.24 \cdot 10^{-28}$ for $y_n$ = $y_{thres}$ = 7.5. This value clearly shows that models using only the radial coordinate (degree of freedom I) are not suitable for studying flip-outs (degree of freedom III).

The requirements for the thresholds for registering open states are completely different in the case of radial separation of DNA strands. The rigidity of the sugar-phosphate backbone and the close coupling of the radial (I) and torsional (II) degrees of freedom (see Figure 2) create local cooperativity of radial openings, see above. As a result, large radial-torsional fluctuations lead to the breakage of complementary H-bonds in several adjacent base pairs at once.

Therefore, only a separation of <u>several adjacent $M$ base pairs</u> above the threshold ($y > y_{thres}$) is considered to be the true open state in numerous investigations of the PBD model [49, 50, 60]. Really, a moderate separation of 3-4 Å is sufficient to accept an H$_2$O molecule between the strands. However, the breakage of complementary H-bonds precisely in $M$ adjacent base pairs is necessary.

Radial-torsional separation of DNA strands (radial opening) can occur at temperatures much lower than its melting point [49, 61, 62]. Local separation of complementary DNA strands is the result of rare high-amplitude fluctuations leading to a concerted displacement of 6-10 base pairs. This follows from numerous experimental data. For example, this is the scale of the "critical length" phenomenon described above [25]. Another reference point is the size of fluorescent labels used to study radial opening of the duplex [61], see Figure 5.3 in [1]. The close relationship between the degrees of freedom (I) and (II) leads to "unwinding" of the Watson-Crick helix at the site of its radial opening. This results in torsional stresses in adjacent DNA regions, see Figure 3(b). The stresses cause the closing of the open region.

The processes described above cannot be considered equilibrium. The locally unwinded region of the Watson-Crick helix is an unstable change. It has a lifetime from several µs to 1 ms [61, 62]. At the same time, "bubble" is the established term for this open state in the literature, see [2, 48, 51, 53, 54, 55, 60, 63]. The analogy with the typical "high-temperature" denaturation bubble, which is the nucleus of a new phase at the onset of a helix-coil transition in DNA, creates another kind of disorder. However, the disorder in this case concerns thermodynamics, but not stereometry.

We estimated the probability of the formation of a stable denatured DNA region in the PBD model in [58]. This probability turned out to be very low, on the order of $10^{-9}$. Furthermore, this probability remains low even near the melting temperature. Our most recent analysis of the thermodynamic properties of the PBD model using computational experiments confirmed that DNA denaturation corresponds to a first-order phase transition [59]. It is well known that during a first-order phase transition, high-amplitude fluctuations in an ideal homogeneous chain are absent.id eros vel

At the same time, the formation of nuclei and even entire regions of a different phase is possible in a non-ideal DNA chain. The denatured regions are stable near a critical temperature. They can exist in the DNA double helix for a very long time. It may seem that the existence of phase transitions in DNA as a quasi-one-dimensional system contradicts general physical considerations. In particular, it contradicts the Landau-Lifshitz assertion that phase transitions are impossible in one-dimensional systems [64]. Furthermore, a phase transition in DNA contradicts the Mermin-Wagner theorem regarding the impossibility of phase transitions in such systems [65].

However, these statements are only valid for continuous systems and rely on certain assumptions about their spectra. The PBD model is not a continuous mechanical model of DNA. It is a semi-continuous (or semi-discrete) model. The spectrum of the PBD model contains a gap that stabilizes the system against long-wavelength fluctuations that denature the molecule. It is also worth remembering that the PBD model was originally developed to study the denaturation behavior of DNA at high temperatures [4].

A set of simple conclusions can be drawn on the base of all of the above.

1. Transient radial separation of DNA strands usually does not lead to the formation of a stable denatured region – a denaturation bubble. The only exceptions are cases of bubble stabilization due to certain factors: e. g., high temperatures, specific DNA-protein interactions, and short DNA oligomer lengths [66]. The latter makes it possible to relieve torsional stresses at the ends of the duplex. See also Fig. 3 in [15].

2. Hence, the real (stable) denaturation bubble either does not form at all or strongly destabilizes a large section of the DNA helix with its formation.

First, the corresponding behavior of the duplex is clearly indicated by the phenomenon of local cooperativity described above. Second, similar behavior is evidenced by the jump in heat capacity obtained in our recent stochastic studies of the PBD model for homogeneous DNA [59].

Of course, the PBD model, like other simple DNA models, is merely a "one-dimensional projection" of a highly complex biomacromolecule that exists in three dimensions. Therefore, no simple model can describe all aspects of the behavior of real DNA. And this is precisely why correct interpretation of experiments is necessary – figuratively speaking, they also need to be "projected" onto one-dimensional models.

The result of this "projection" is obvious. Not all changes of coordinates in the PBD model can be correlated strictly with the DNA open states. For example, the detection of a real radial separation of the DNA strands at $y > y_{thres} = 3$ Å in $M = 10$ adjacent base pairs is an example of the correct approach [60]. The value of the equilibrium constant $K_d$ obtained in the cited study for the given parameters is quite plausible – on the order of $10^{-5}$. This value accurately reflects the short-term nature of the local radial opening. Anything associated with smaller values of M and/or ythres can only reflect "subthreshold" conformational changes in DNA. These changes are not open states.

A striking example is the aforementioned case of β-premeltons (see Figure 5). Traditional methods for studying open states do not allow their detection, since such conformational changes are not associated with the breakage of complementary H-bonds. Nevertheless, estimates of the thermodynamic properties of analogous conformational changes can be found in the literature — and they are quite interesting. For example, in the paper of B. Bouvier and H. Grubmuller, there are indications of a very high concentration of conformational rearrangements detected by NMR methods, but incapable of proton exchange [36]. According to the authors, this state is "2–3 kcal/mol higher in energy than the closed state" and "separated from the latter by a barrier of ~16 kcal/mol." Estimating the equilibrium kinetics at such $\Delta G°$ values suggests very high concentrations of the mysterious state — up to 0.03 [1]. Most likely, this conformational change corresponds specifically to β-premelton.

The NMR data indicate that the energy of nucleotide oscillations can be "discharged" into the potential energy of various conformational rearrangements. However, not all conformational changes lead to DNA opening. Therefore, the threshold $y_{thres}$ should be replaced by an "effective threshold coordinate" when comparing the behavior of the PBD model with experiments. This coordinate has the form $[M, y_{thres}]$, where the number of opened adjacent base pairs is $M \geq 8$. Radial separation of DNA strands corresponds to the open state only when the conditions $M \geq 8$ and $y_n \geq y_{thres}$ are met for all adjacent base pairs; otherwise, some "subthreshold" conformational change occurs.

The analysis held in the current investigation can also serve as a guide for reparametrizing the PBD model or any other. For example, the parameters by Campa and Giansanti [57], traditionally used in most PBD model studies, have been shown to allow, among other things, reproducing the effect of the critical oligomer length [63]. Hence, at $k = 0.025$ eV Å$^{-2}$ and $\rho = 2$ (see above), the PBD model is fully capable of accounting for local cooperativity in DNA. At the same time, there is an alternative parameterization of the PBD model: $k = 0.00045$ eV·Å$^{-2}$, $\rho = 50$, $\alpha = 0.2$ Å$^{-1}$, $D_{AT} = 0.1255$ eV, $D_{GC} = 0.1655$ eV, $a_{AT} = 4.2$ Å$^{-1}$, $a_{GC} = 6.9$ Å$^{-1}$ [56, 67]. The extent of local cooperativity in the PBD model with such a small value of the elastic constant $k$ requires additional verification.

In our opinion, no radical reparameterizations are required for improvement of the agreement between the PBD model and experiments. A moderate adjustment of the key parameters – the wells' depths of the on-site potentials DAT and DGC, the elastic constant $k$, and the cooperativity parameter ρ — in the range of 20-35% is sufficient to optimize the PBD model. Even small changes in these parameters make it is possible to achieve a virtually exact agreement between theory and experiment within the framework of the concept of an effective total threshold coordinate.

**CONCLUSION**

The present study addresses the issue of degrees of freedom of bases in simple mechanical models of DNA. The development and parameterization of any model is based on experiments. At the same time, different experiments are devoted to the study of different types of the DNA open state. The formation of different types of open state involves different degrees of freedom of the bases. A typical simple mechanical model of DNA includes an incomplete set of degrees of freedom and, therefore, is generally capable of describing only one type of DNA open state. Therefore, the precise correspondence between the types of the DNA open state in the model and the experiment is one of the key points of modeling. Matching errors can lead to the loss of the main advantages of any DNA model.

We made use of the Peyard-Bishop-Dauxois (PBD) mechanical model as an example [4]. This model uses the radial degree of freedom of base pairs to describe the duplex's state. This degree of freedom is closely related to the torsional stresses of the sugar-phosphate backbone in real DNA. Therefore, the radial openings of several adjacent base pairs occur in a cooperative manner. We present experimental evidences for this cooperativity. Based on the analysis of experiments, a new universal criterion for the open state in those DNA models where the radial degree of freedom of the base pair is taken into account is proposed.

Comparisons of the probabilities of radial openings in the PBD model with the experimental kinetics of single-base angular openings are shown to be unacceptable. Furthermore, the threshold values for radial "opening" of single base pairs in most PBD model investigations are too low. Undoubtedly, the "opening" kinetics obtained in many theoretical investigations are important for understanding the kinetics of conformational transitions preceding the formation of open states. We provide an example of one such transition. Nevertheless, these kinetics are not directly related to the open states of DNA. In our opinion, the best approach to studying the DNA radial openings is to introduce an effective total threshold coordinate [$M$, $y_{thres}$] for the radial displacement of base pairs.